\documentclass[12pt,a4paper]{article}

\usepackage{lineno}  

\usepackage{graphicx}  

\usepackage{xspace}
\usepackage{color}
\usepackage{colortbl}

\usepackage{amsmath}

\usepackage{ifthen} 

\newboolean{pdflatex}
\setboolean{pdflatex}{true} 
\newboolean{articletitles}
\setboolean{articletitles}{true} 

\newboolean{uprightparticles}
\setboolean{uprightparticles}{false} 
\usepackage{amssymb}
\usepackage{amsfonts}
\usepackage{upgreek}

\usepackage{hyperref}
\usepackage[all]{hypcap}

\def\lhcb {LHCb\xspace}
\def\ux85 {UX85\xspace}

\def\dzero  {D\O\xspace}

\ifthenelse{\boolean{uprightparticles}}%
{

 \def\Ppsi        {\ensuremath{\uppsi}\xspace}

 \def\PDelta      {\ensuremath{\Delta}\xspace}                 
 \def\PXi      {\ensuremath{\Xi}\xspace}                 
 \def\PLambda      {\ensuremath{\Lambda}\xspace}                 
 \def\PSigma      {\ensuremath{\Sigma}\xspace}                 
 \def\POmega      {\ensuremath{\Omega}\xspace}                 
 \def\PUpsilon      {\ensuremath{\Upsilon}\xspace}                 
 
 \def\PB      {\ensuremath{\mathrm{B}}\xspace}                 
                  
 \def\PD      {\ensuremath{\mathrm{D}}\xspace}

 \def\PJ      {\ensuremath{\mathrm{J}}\xspace}                 
 \def\PK      {\ensuremath{\mathrm{K}}\xspace}

 \def\Pb      {\ensuremath{\mathrm{b}}\xspace}                 
 \def\Pc      {\ensuremath{\mathrm{c}}\xspace}

 \def\Pi      {\ensuremath{\mathrm{i}}\xspace}

}
{

 \def\Ppsi        {\ensuremath{\psi}\xspace}                 
                  
 \mathchardef\PDelta="7101
 \mathchardef\PXi="7104
 \mathchardef\PLambda="7103
 \mathchardef\PSigma="7106
 \mathchardef\POmega="710A
 \mathchardef\PUpsilon="7107
                  
 \def\PB      {\ensuremath{B}\xspace}                 
                  
 \def\PD      {\ensuremath{D}\xspace}

 \def\PJ      {\ensuremath{J}\xspace}                 
 \def\PK      {\ensuremath{K}\xspace}

 \def\Pb      {\ensuremath{b}\xspace}                 
 \def\Pc      {\ensuremath{c}\xspace}

 \def\Pi      {\ensuremath{i}\xspace}

}

\def\cquark    {\ensuremath{\Pc}\xspace}

\def\bquark    {\ensuremath{\Pb}\xspace}

\def\kaon  {\ensuremath{\PK}\xspace}
  \def\Kbar  {\kern 0.2em\overline{\kern -0.2em \PK}{}\xspace}

\def\Kz    {\ensuremath{\kaon^0}\xspace}
\def\Kzb   {\ensuremath{\Kbar^0}\xspace}
\def\KzKzb {\ensuremath{\Kz \kern -0.16em \Kzb}\xspace}
\def\Kp    {\ensuremath{\kaon^+}\xspace}
\def\Km    {\ensuremath{\kaon^-}\xspace}

\def\KpKm  {\ensuremath{\Kp \kern -0.16em \Km}\xspace}

  \def\Dbar    {\kern 0.2em\overline{\kern -0.2em \PD}{}\xspace}
\def\D       {\ensuremath{\PD}\xspace}

\def\Dz      {\ensuremath{\D^0}\xspace}
\def\Dzb     {\ensuremath{\Dbar^0}\xspace}
\def\DzDzb   {\ensuremath{\Dz {\kern -0.16em \Dzb}}\xspace}
\def\Dp      {\ensuremath{\D^+}\xspace}
\def\Dm      {\ensuremath{\D^-}\xspace}

\def\DpDm    {\ensuremath{\Dp {\kern -0.16em \Dm}}\xspace}

  \def\Bbar    {\kern 0.18em\overline{\kern -0.18em \PB}{}\xspace}

\def\jpsi     {\ensuremath{{\PJ\mskip -3mu/\mskip -2mu\Ppsi\mskip 2mu}}\xspace}

  \def\Y#1S{\ensuremath{\PUpsilon{(#1S)}}\xspace}

\def\Lbar {\ensuremath{\kern 0.1em\overline{\kern -0.1em\Lambda\kern -0.05em}\kern 0.05em{}}\xspace}

\def\BR         {\BF}

\def\to                 {\ensuremath{\rightarrow}\xspace}

\def\AT#1     {\ensuremath{A_{\mathrm{T}}^{#1}}\xspace}           

\def\C#1      {\ensuremath{\mathcal{C}_{#1}}\xspace}                       
\def\Cp#1     {\ensuremath{\mathcal{C}_{#1}^{'}}\xspace}                    
\def\Ceff#1   {\ensuremath{\mathcal{C}_{#1}^{\mathrm{(eff)}}}\xspace}        
\def\Cpeff#1  {\ensuremath{\mathcal{C}_{#1}^{'\mathrm{(eff)}}}\xspace}       
\def\Ope#1    {\ensuremath{\mathcal{O}_{#1}}\xspace}                       
\def\Opep#1   {\ensuremath{\mathcal{O}_{#1}^{'}}\xspace}                    

\newcommand{\tev}{\ensuremath{\mathrm{\,Te\kern -0.1em V}}\xspace}
\newcommand{\gev}{\ensuremath{\mathrm{\,Ge\kern -0.1em V}}\xspace}
\newcommand{\mev}{\ensuremath{\mathrm{\,Me\kern -0.1em V}}\xspace}
\newcommand{\kev}{\ensuremath{\mathrm{\,ke\kern -0.1em V}}\xspace}
\newcommand{\ev}{\ensuremath{\mathrm{\,e\kern -0.1em V}}\xspace}
\newcommand{\gevc}{\ensuremath{{\mathrm{\,Ge\kern -0.1em V\!/}c}}\xspace}
\newcommand{\mevc}{\ensuremath{{\mathrm{\,Me\kern -0.1em V\!/}c}}\xspace}
\newcommand{\gevcc}{\ensuremath{{\mathrm{\,Ge\kern -0.1em V\!/}c^2}}\xspace}
\newcommand{\gevgevcccc}{\ensuremath{{\mathrm{\,Ge\kern -0.1em V^2\!/}c^4}}\xspace}
\newcommand{\mevcc}{\ensuremath{{\mathrm{\,Me\kern -0.1em V\!/}c^2}}\xspace}

\def\mum  {\ensuremath{\,\upmu\rm m}\xspace}

\def\gsim{{~\raise.15em\hbox{$>$}\kern-.85em
          \lower.35em\hbox{$\sim$}~}\xspace}
\def\lsim{{~\raise.15em\hbox{$<$}\kern-.85em
          \lower.35em\hbox{$\sim$}~}\xspace}

\def\PDF {PDF\xspace}

\def\tell1  {TELL1\xspace}
\def\ukl1   {UKL1\xspace}

\usepackage{cite}
\usepackage{mciteplus}
\def\bcjppp{B_c^+\to\jpsi\pi^+\pi^-\pi^+}
\def\bcjp{B_c^+\to\jpsi\pi^+}
\def\bcjpopp{B_c^+\to\jpsi\pi^+[\pi^-\pi^+]}
\def\bujppp{B^+\to\jpsi\pi^+\pi^-\pi^+}
\def\bujkpp{B^+\to\jpsi K^+\pi^-\pi^+}
\def\bujk{B^+\to\jpsi K^+}

\def\BR{{\cal B}}
\def\DLL{{\rm DLL}}
\def\PDF{{\cal P}}
\def\NDOF{\hbox{\rm NDOF}}
\def\cospj{\cos(\pi,\jpsi)}

\def\r31{R_{3/1}}

\begin{document}

\renewcommand{\thefootnote}{\fnsymbol{footnote}}
\setcounter{footnote}{1}


\begin{titlepage}
\pagenumbering{roman}

\vspace*{-1.5cm}
\centerline{\large EUROPEAN ORGANIZATION FOR NUCLEAR RESEARCH (CERN)}
\vspace*{1.5cm}
\hspace*{-0.5cm}
\begin{tabular*}{\linewidth}{lc@{\extracolsep{\fill}}r}
\ifthenelse{\boolean{pdflatex}}
{\vspace*{-2.7cm}\mbox{\!\!\!\includegraphics[width=.14\textwidth]{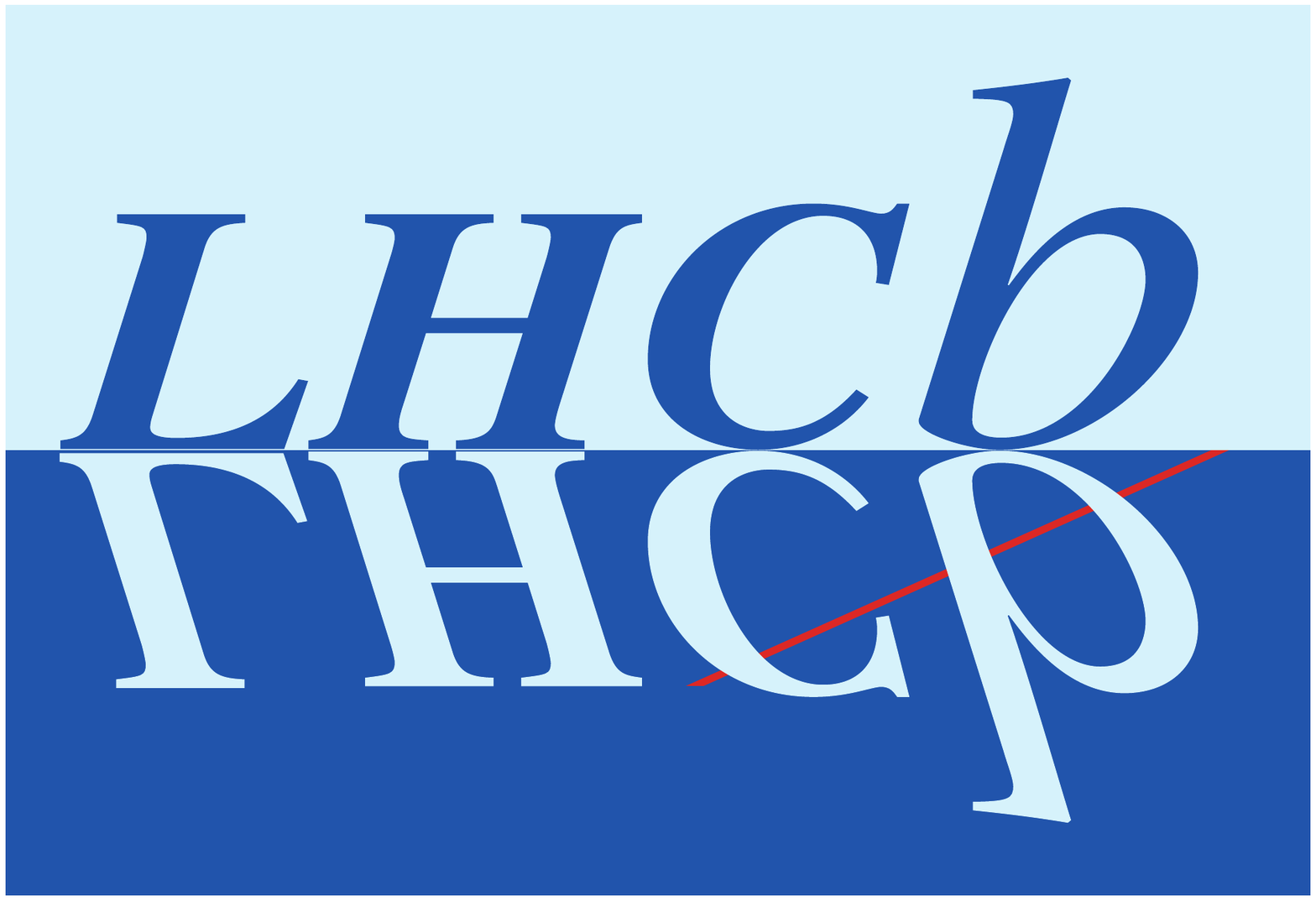}} & &}%
{\vspace*{-1.2cm}\mbox{\!\!\!\includegraphics[width=.12\textwidth]{lhcb-logo.eps}} & &}%
\\
 & & CERN-PH-EP-2012-090 \\  
 & & LHCb-PAPER-2011-044 \\  
 & & 31 March 2012 \\ 
 & & \\
\end{tabular*}

\vspace*{4.0cm}

{\bf\boldmath\huge
\begin{center}
  First observation of the decay $B_c^+\to J/\psi \pi^+\pi^-\pi^+$
\end{center}
}

\vspace*{2.0cm}

\begin{center}
The LHCb collaboration
\footnote{Authors are listed on the following pages.}
\end{center}

\vspace{\fill}

\begin{abstract}
  \noindent
The decay $B_c^+\to J/\psi \pi^+\pi^-\pi^+$ is observed for the first time,
using 0.8 fb$^{-1}$ of $pp$ collisions at $\sqrt{s}=7$ TeV collected by the LHCb experiment.
The ratio of branching fractions
${\cal B}(B_c^+\to J/\psi \pi^+\pi^-\pi^+)/{\cal B}(B_c^+\to J/\psi \pi^+)$
is measured to be $2.41\pm0.30\pm0.33$,
where the first uncertainty is statistical and the second systematic. 
The result is in agreement with theoretical predictions. 

\end{abstract}

\vspace*{2.0cm}

\begin{center}
  Submitted to Physical Review Letters
\end{center}

\vspace{\fill}

\end{titlepage}


\newpage
\setcounter{page}{2}
\mbox{~}
\newpage

\centerline{\large\bf LHCb collaboration}
\begin{flushleft}
\small
R.~Aaij$^{38}$, 
C.~Abellan~Beteta$^{33,n}$, 
B.~Adeva$^{34}$, 
M.~Adinolfi$^{43}$, 
C.~Adrover$^{6}$, 
A.~Affolder$^{49}$, 
Z.~Ajaltouni$^{5}$, 
J.~Albrecht$^{35}$, 
F.~Alessio$^{35}$, 
M.~Alexander$^{48}$
% ...

{\footnotesize \it
$ ^{1}$Centro Brasileiro de Pesquisas F\'{i}sicas (CBPF), Rio de Janeiro, Brazil\\
% ...
$ ^{5}$Clermont Universit\'{e}, Universit\'{e} Blaise Pascal, CNRS/IN2P3, LPC, Clermont-Ferrand, France\\
$ ^{6}$CPPM, Aix-Marseille Universit\'{e}, CNRS/IN2P3, Marseille, France\\
% ...
$ ^{33}$Universitat de Barcelona, Barcelona, Spain\\
$ ^{34}$Universidad de Santiago de Compostela, Santiago de Compostela, Spain\\
$ ^{35}$European Organization for Nuclear Research (CERN), Geneva, Switzerland\\
% ...
$ ^{38}$Nikhef National Institute for Subatomic Physics, Amsterdam, The Netherlands\\
% ...
$ ^{43}$H.H. Wills Physics Laboratory, University of Bristol, Bristol, United Kingdom\\
% ...
$ ^{48}$School of Physics and Astronomy, University of Glasgow, Glasgow, United Kingdom\\
$ ^{49}$Oliver Lodge Laboratory, University of Liverpool, Liverpool, United Kingdom\\
% ...
$ ^{n}$LIFAELS, La Salle, Universitat Ramon Llull, Barcelona, Spain\\
% ...
}
\end{flushleft}

\cleardoublepage

\renewcommand{\thefootnote}{\arabic{footnote}}
\setcounter{footnote}{0}



\pagestyle{plain} 
\setcounter{page}{1}
\pagenumbering{arabic}


%

\newlength{\figsize}
\setlength{\figsize}{0.7\hsize}
\def\bcjpopp{B_c^+\to\jpsi\pi^+[\pi^-\pi^+]}
\def\jpopp{\jpsi\pi^+[\pi^-\pi^+]}
\def\bcjppp{B_c^+\to\jpsi\pi^+\pi^-\pi^+}
\def\bc2sp{B_c^+\to\psi(2S)\pi^+}
\def\bu2sp{B^+\to\psi(2S)\pi^+}
\def\p2sppj{\psi(2S)\to\pi^+\pi^-\jpsi}
\def\bcjp{B_c^+\to\jpsi\pi^+}
\def\bujp{B^+\to\jpsi\pi^+}
\def\budppp{B^+\to \bar{D}^{*0} \pi^+\pi^-\pi^+}
\def\budp{B^+\to \bar{D}^{*0} \pi^+}
\def\bujppp{B^+\to\jpsi\pi^+\pi^-\pi^+}
\def\bujkpp{B^+\to\jpsi K^+\pi^-\pi^+}
\def\bujk{B^+\to\jpsi K^+}
\def\cospj{\cos(\pi,\jpsi)}
\def\r31{R_{3/1}}
\def\BR{{\cal B}}
\def\DLL{{\rm DLL}}
\def\PDF{{\cal P}}
\def\NDOF{\hbox{\rm ndf}}
\def\R31{\BR(\bcjppp)/\BR(\bcjp)}

\noindent
The $B_c^+$ meson is the ground state of the $\bar{b}c$ quark pair system\footnote{Charge-conjugate states are implied in this Letter.}. 
Studies of its properties are important, since it is the only meson consisting of two different heavy quarks. 
It is also the only meson in which decays of both constituents compete with each other. 
Numerous predictions for $B_c^+$ branching fractions have been published (for a review see e.g.~Ref.~\cite{Brambilla:2004wf}).
To date, no measurements exist which would allow to test these predictions, even in ratios. 
$B_c^+$ production rates are about three orders of magnitude smaller at high energy colliders than for 
the other $B$ mesons composed of a $b$ quark and a light quark ($B^+$, $B^0$ and $B_s^0$). 
On the experimental side, whatever is known about the $B_c^+$ meson was measured at the Tevatron. 
It was discovered by the CDF experiment in the semileptonic decay, $B_c^+\to \jpsi l^+\nu X$ \cite{Abe:1998wi}.
This decay mode was later used to measure the $B_c^+$ lifetime \cite{Abulencia:2006zu,Abazov:2008rba},
which is a factor of three shorter than for the other $B$ mesons
as both $b$ and $c$ quark may decay.
Only one hadronic decay mode of $B_c^+$ has been observed so far, $\bcjp$. 
It was utilized by CDF \cite{Aaltonen:2007gv} and \dzero \cite{Abazov:2008kv}
to measure the $B_c^+$ mass\footnote{We use mass and momentum units in which $c=1$.}, 
$6277\pm6$ MeV \cite{PDG}.

In this Letter, the first observation of the decay mode $\bcjppp$ is presented 
using a data sample corresponding to an integrated luminosity of $0.8$~fb$^{-1}$  
collected in 2011 by the LHCb detector \cite{Alves:2008zz}, in $pp$ collisions at the LHC at $\sqrt{s}=7$~TeV.
The branching fraction for this decay is expected to be $1.5-2.3$ times
higher than for $\bcjp$ \cite{Rakitin:2009ya,Likhoded:2009ib}.
However, the larger number of pions in the final state results in
a smaller total detection efficiency due to the limited detector 
acceptance.
We measure the $\bcjppp$ branching fraction relative to that for the $\bcjp$ decay 
and test the above theoretical predictions.

The \lhcb detector~\cite{Alves:2008zz} is a single-arm forward
spectrometer covering the pseudo-rapidity range $2<\eta <5$, designed
for the study of particles containing \bquark or \cquark quarks. The
detector includes a high precision tracking system consisting of a
silicon-strip vertex detector surrounding the $pp$ interaction region,
a large-area silicon-strip detector located upstream of a dipole
magnet with a bending power of about $4{\rm\,Tm}$, and three stations
of silicon-strip detectors and straw drift-tubes placed
downstream. The combined tracking system has a momentum resolution
$\Delta p/p$ that varies from 0.4\% at 5~GeV to 0.6\% at 100~GeV,
and an impact parameter (IP) resolution of 20\mum for tracks with high
transverse momentum. Charged hadrons are identified using two
ring-imaging Cherenkov detectors. Photon, electron and hadron
candidates are identified by a calorimeter system consisting of
scintillating-pad and pre-shower detectors, an electromagnetic
calorimeter and a hadronic calorimeter. Muons are identified by a muon
system composed of alternating layers of iron and multiwire
proportional chambers. 
The muon system, electromagnetic and hadron calorimeters
provide the capability of first-level hardware triggering.
The single and dimuon hardware triggers provide good efficiency 
for $\bcjpopp$, $\jpsi\to\mu^+\mu^-$ events. 
Here, $\pi^+[\pi^-\pi^+]$ stands for either $\pi^+$ or $\pi^+\pi^-\pi^+$ depending on the
$B_c^+$ decay mode. 
Events passing the hardware trigger are read out and sent to an event-filter farm for further processing. 
Here, a software-based two-stage trigger reduces the rate from 1 MHz to about 3 kHz.
The most efficient software triggers \cite{LHCb-PUB-2011-016} 
for this analysis require a charged track with transverse momentum ($p_{\rm T}$) of more than $1.7$~GeV 
($p_{\rm T}>1.0$~GeV if identified as muon) 
and with an IP to any primary $pp$-interaction vertex (PV) larger than $100$~$\mu$m.
A dimuon trigger requiring $p_{\rm T}(\mu)>0.5$~GeV, large dimuon mass, $M(\mu^+\mu^-)>2.7$~GeV,
and with no IP requirement complements the single track triggers.
At the final stage, we either 
require a $\jpsi\to\mu^+\mu^-$ candidate with $p_{\rm T}>2.7$~GeV ($>1.5$~GeV in the first 42\%\ of data)
or a muon-track pair with significant IP.
  
In the subsequent offline analysis of the data, 
$\jpsi\to\mu^+\mu^-$ candidates are selected with the following criteria: $p_{\rm T}(\mu)>0.9$~GeV,
$p_{\rm T}(\jpsi)>3.0$~GeV ($>1.5$~GeV in the first 42\%\ of data),   
$\chi^2$ per degree of freedom of the two muons forming a common vertex, $\chi^2_{\rm vtx}(\mu^+\mu^-)/\NDOF<9$,
and a mass window $3.04< M(\mu^+\mu^-)<3.14$~GeV.   
We then find 
$\pi^+\pi^-\pi^+$ combinations consistent with originating from a common vertex with 
$\chi^2_{\rm vtx}(\pi^+\pi^-\pi^+)/\NDOF<9$, 
with each pion separated from all PVs by at least three standard deviations ($\chi^2_{\rm IP}(\pi)>9$),
and having $p_{\rm T}(\pi)>0.25$~GeV.
A loose kaon veto is applied using the particle identification system.
A five-track $\jpsi \pi^+\pi^-\pi^+$ vertex is formed ($\chi^2_{\rm vtx}(\jpsi \pi^+\pi^-\pi^+)/\NDOF<9$).
To look for candidates in the normalization mode, $\bcjp$, the criteria $p_{\rm T}(\pi)>1.5$~GeV and
$\chi^2_{\rm vtx}(\jpsi \pi^+)/\NDOF<16$ are used. 
All $B_c^+$ candidates are required to have $p_{\rm T}>4.0$~GeV and a decay time 
of at least $0.25$~ps.
When more than one PV is reconstructed, 
that which gives the smallest IP significance for the $B_c^+$ candidate is chosen.
The invariant mass of a $\mu^+\mu^- \pi^+[\pi^-\pi^+]$ combination is evaluated after 
the muon pair is constrained to the $\jpsi$ mass and all final state particles are constrained to form 
a common vertex. 

\begin{figure}[t]
  \begin{center}
  \ifthenelse{\boolean{pdflatex}}{
    \includegraphics*[width=\figsize]{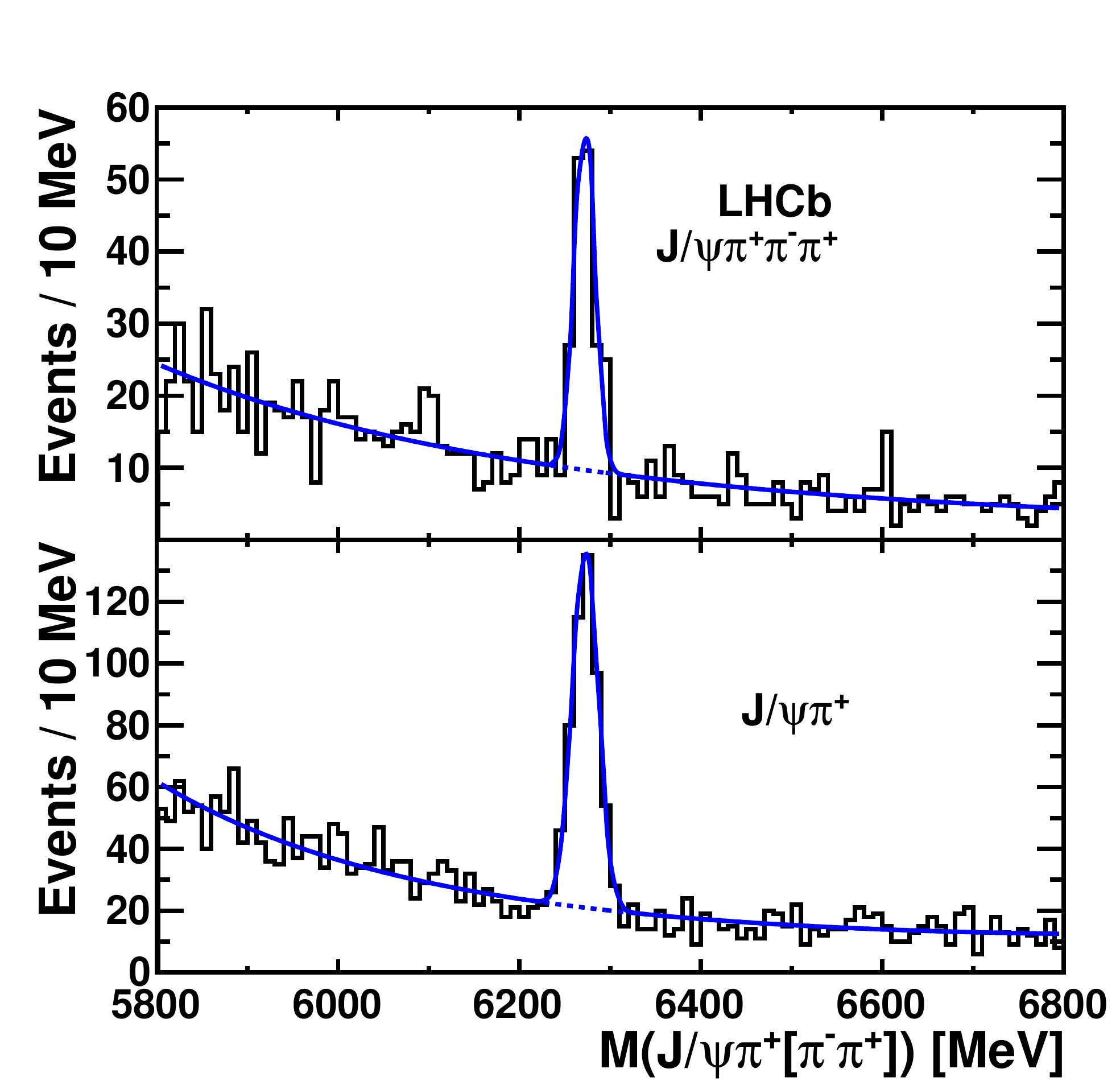}
   }{
    \includegraphics*[width=\figsize]{fig3p_bcfits.eps}
   } 
  \end{center}
  \vskip-0.3cm\caption{\small 
    Invariant mass distribution of $\bcjppp$ (top) and $\bcjp$ (bottom) candidates.   
    The maximum likelihood fits of $B_c^+$ signals are superimposed.}
  \label{fig:bcfits}
\end{figure}

Further background suppression 
is provided by an event selection based on a likelihood ratio. 
In the case of uncorrelated input variables this provides 
the most efficient discrimination between signal and background. 
The overall likelihood is a product of 
the probability density functions (PDFs), $\PDF(x_i)$,  
for the four sensitive variables ($x_i$): 
smallest $\chi^2_{\rm IP}(\pi)$ among the pion candidates,
$\chi^2_{\rm vtx}(\jpsi  \pi^+[\pi^-\pi^+])/\NDOF$,
$B_c^+$ candidate IP significance, $\chi^2_{\rm IP}(B_c)$,
and cosine of the largest opening angle between the $\jpsi$ and pion candidates 
in the plane transverse to the beam.
The latter peaks at positive values for the signal as the $B_c^+$ meson has a high transverse momentum.
Background events that combine particles from two different $B$ mesons peak at negative values,
whilst background events that include random combinations of tracks are uniformly distributed.  
The signal PDFs, $\PDF_{\rm sig}(x_i)$, 
are obtained from a Monte Carlo simulation of 
$\bcjpopp$ decays.
The background PDFs, $\PDF_{\rm bkg}(x_i)$, are obtained from the data 
with a $\jpsi \pi^+[\pi^-\pi^+]$ invariant mass in the range $5.35-5.80$~GeV or $6.80-8.50$~GeV (far-sidebands).
A logarithm of the ratio of the signal and background PDFs is formed:
$\DLL_{\rm sig/bkg} = -2 \sum_{i=1}^4 \ln(\PDF_{\rm sig}(x_i)/\PDF_{\rm bkg}(x_i))$.
Requirements on the log-likelihood ratio, $\DLL_{\rm sig/bkg}<-5$ for $\bcjppp$ and $\DLL_{\rm sig/bkg}<-1$ for $\bcjp$, 
have been chosen to  
maximize $N_{\rm sig}/\sqrt{N_{\rm sig}+N_{\rm bkg}}$, 
where $N_{\rm sig}$ is the expected $\bcjpopp$ signal yield and
the $N_{\rm bkg}$ is the background yield in the $B_c^+$ peak region ($\pm2.5\,\sigma$).
The absolute normalization of $N_{\rm sig}$ and $N_{\rm bkg}$ is obtained from a fit 
to the $\jpopp$ invariant mass distribution
with $\DLL_{\rm sig/bkg}<0$, while their dependence on 
the $\DLL_{\rm sig/bkg}$ requirement comes from the signal simulation 
and the far-sidebands, respectively.  
The $\jpopp$ mass distributions after applying all requirements are shown in Fig.~\ref{fig:bcfits}.
To determine the signal yields,
a Gaussian signal shape with position and width as free parameters 
is fitted to these distributions 
on top of a background assumed to be
an exponential function with a second order polynomial as argument.
We observe $135\pm14$ $\bcjppp$ and $414\pm25$ $\bcjp$ signal events. 
Using different signal and background parameterizations in the 
fits, the ratio of the signal yields changes by up to 3\%. 
  
\begin{figure}[t]
  \begin{center}
  \ifthenelse{\boolean{pdflatex}}{
    \includegraphics*[width=\figsize]{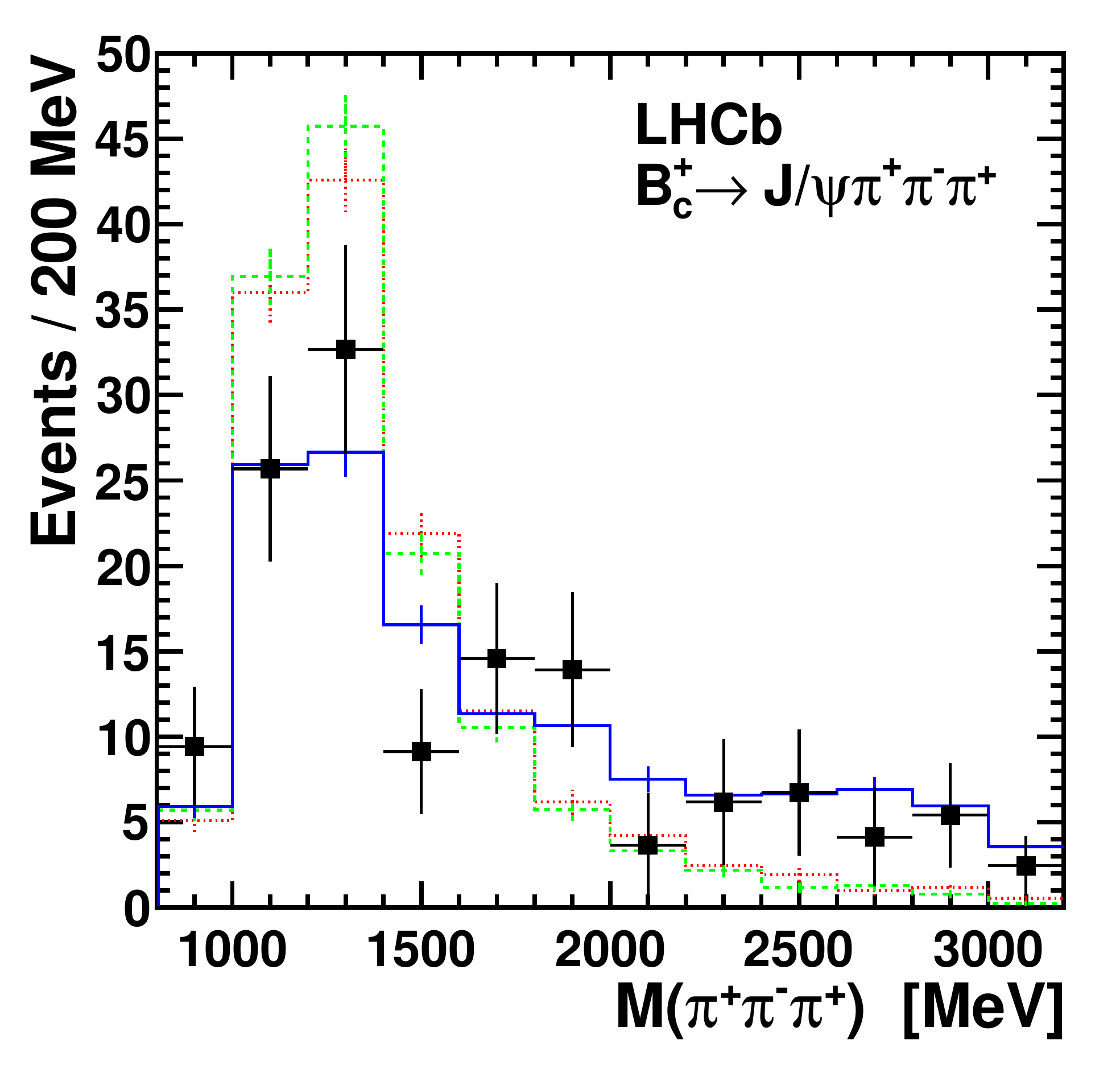}
   }{
    \includegraphics*[width=\figsize]{fig3p_ppp.eps}
   } 
  \end{center}
  \vskip-0.3cm\caption{\small
    Invariant mass distribution of the $\pi^+\pi^-\pi^+$ combinations
    for the sideband-subtracted $\bcjppp$ data (points) 
    and signal simulation (lines).
    The solid blue line corresponds to the BLL simulations, the PH model is shown as a green dashed line and
    the PHPOL model is shown as a red dotted line.
    All error bars are statistical.
    }
  \label{fig:mppp}
\end{figure}

\begin{figure}[t]
  \begin{center}
  \ifthenelse{\boolean{pdflatex}}{
    \includegraphics*[width=\figsize]{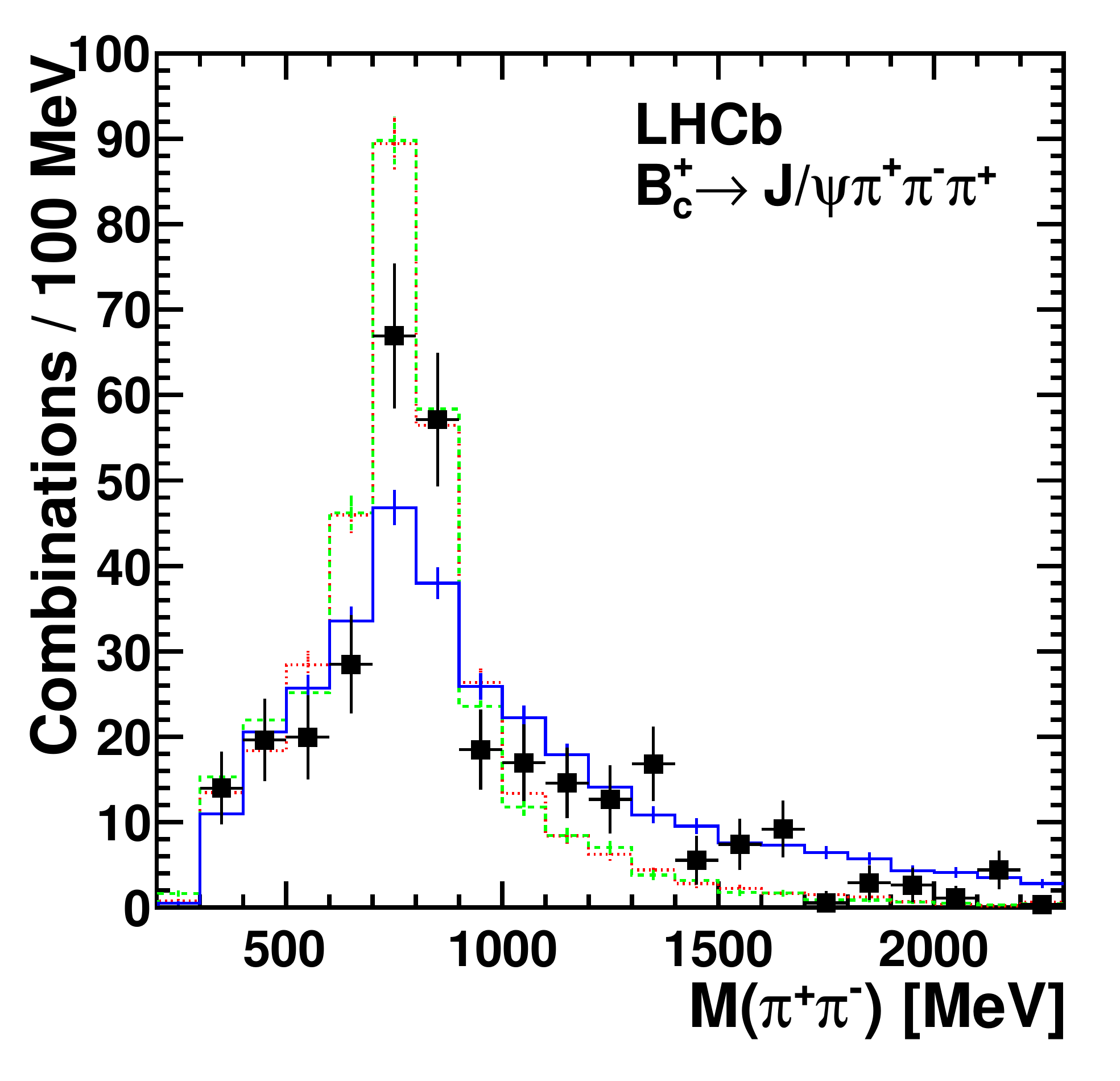}
   }{
    \includegraphics*[width=\figsize]{fig3p_pp.eps}
   } 
  \end{center}
  \vskip-0.3cm\caption{\small 
    Invariant mass distribution of the $\pi^+\pi^-$ combinations
    (two entries per $B_c^+$ candidate) for the 
    sideband-subtracted $\bcjppp$ data (points) 
    and signal simulation (lines).
    The solid blue line corresponds to the BLL simulations, the PH model is shown as a green dashed line and
    the PHPOL model is shown as a red dotted line.
    All error bars are statistical.
    }
  \label{fig:mpp}
\end{figure}

\begin{figure}[t]
  \begin{center}
  \ifthenelse{\boolean{pdflatex}}{
    \includegraphics*[width=\figsize]{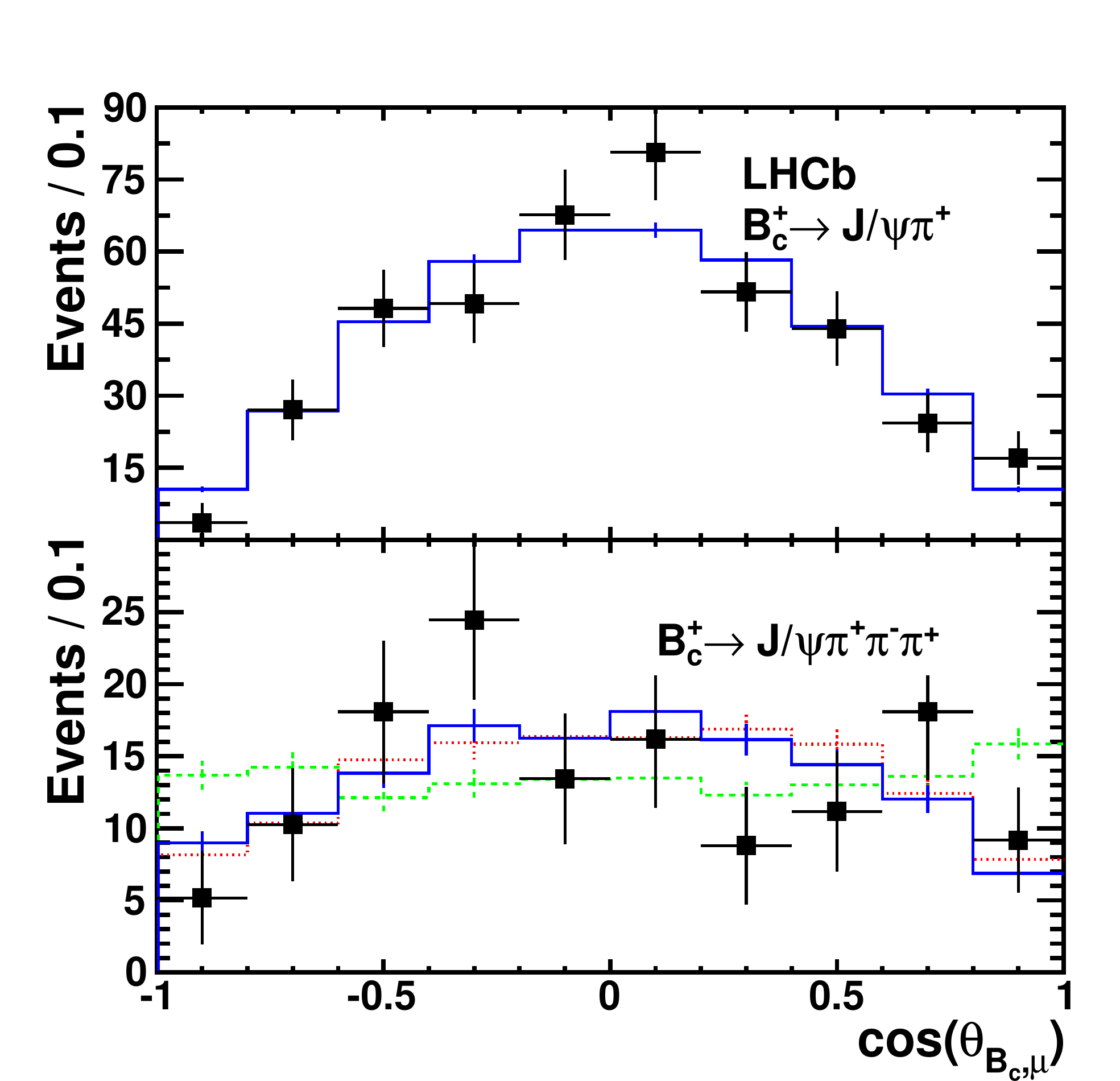}
   }{
    \includegraphics*[width=\figsize]{fig3p_muhel.eps}
   } 
  \end{center}
  \vskip-0.3cm\caption{\small 
    Distributions of the cosine of the angle between the $\mu^+$ and $B_c^+$ boosted to the rest frame of the $\jpsi$ meson 
    for the sideband-subtracted $\bcjp$ (top) and $\bcjppp$ (bottom) data (points) 
    and signal simulation (lines).
    In the bottom plot, the solid blue line corresponds to the BLL simulations, the PH model is shown as a green dashed line and
    the PHPOL model is shown as a red dotted line.
    All error bars are statistical.
    }
  \label{fig:hel}
\end{figure}

The ratio of event yields is converted into a measurement of the ratio of branching fractions $\R31$, where
we rely on the simulation for the determination of the ratio of event selection efficiencies. 
The production of $B_c^+$ mesons is simulated using the BCVEGPY generator \cite{Chang:2003cq,Chang:2006xka}
which gives a good description of the observed transverse momentum and pseudorapidity distributions in our data.
The simulation of the two-body $\bcjp$ decay takes into account the spins of the particles and 
contains no ambiguities.
The phenomenological model by Berezhnoy, Likhoded and Luchinsky \cite{Likhoded:2009ib,Berezhnoy:2011nx} 
(BLL) is used to simulate $\bcjppp$ decays. 
This model, which is based on amplitude factorisation into hadronic and weak currents,
implements $B_c^+\to\jpsi W^{+*}$ axial-vector form-factors
and a $W^{+*}\to\pi^+\pi^-\pi^+$ decay via the exchange of the virtual 
$a_1^+(1260)$ and $\rho^0(770)$ resonances. 
Since it is not possible to identify 
which of the same-sign pions originates from
the $\rho^0$ decay, the two $\rho^0$ paths interfere.
To explore the model dependence of the efficiency we also use two phase-space models, 
implementing the same decay chain with no interference and with either 
no polarization in the decay (PH) 
or helicity amplitudes of 0.46, 0.87 and 0.20 
for $+1$, $0$ and $-1$ $\jpsi$ helicities (PHPOL), respectively.
For the helicity structure in the PHPOL model, 
we use the expectation for the $B^+\to D^{*0} a_1^+(1260)$ decay based 
on QCD factorisation \cite{Rosner:1990xx}.
The background-subtracted distribution\footnote{For comparisons between the data and simulation we use the data 
within $\pm2.5\,\sigma$ of the observed peak position in the $B_c^+$ mass (signal region).
We subtract the background distributions as estimated from the
$\pm(5-30)\,\sigma$ near-sidebands.} 
of the $M(\pi^+\pi^-\pi^+)$ mass for the $\bcjppp$ data shown in Fig.~\ref{fig:mppp}
exhibits an $a_1^+(1260)$ peak and favours the BLL model.
The $\rho^0(770)$ peak in the $M(\pi^+\pi^-)$ mass distribution shown in Fig.~\ref{fig:mpp}
is smaller than in the two phase-space models, 
but more pronounced than in the BLL model, with the tail favouring the BLL model.
The $\jpsi$ helicity angle distribution shown in Fig.~\ref{fig:hel} disfavours the model with no polarization.
Since the BLL model gives the best overall description of the data, we choose it to evaluate 
the central value of the ratio of $\bcjppp$ to $\bcjp$ efficiencies, $0.135\pm0.004$, and
use the phase-space models to quantify the systematic uncertainty.
The phase-space models produce relative efficiencies different by $-9\%$ (PHPOL) and $+5\%$ (PH). 
We assign a 9\%\ systematic uncertainty to the model dependence of $\bcjppp$ efficiency.

The distribution of the $M(\jpsi\pi^+\pi^-)$ mass 
has an isolated peak of four events at the $\psi(2S)$ mass. 
From the $B_c^+$ sidebands we expect $0.50\pm0.25$ background events in this peak.
This is consistent with $3.6\pm0.6$ expected $\bc2sp$ events, 
assuming $\BR(\bc2sp)/\BR(\bcjp)$ equals to $\BR(\bu2sp)/\BR(\bujp)=0.52\pm0.07$ \cite{PDG} 
after subtracting 10\%\ to account for the phase-space difference.
Since this contribution is only $(2.6\pm1.5)\%$ of the $\bcjppp$ signal yield, 
we do not subtract it and assign a $2\%$ systematic uncertainty to the ratio of the branching fractions 
due to the efficiency difference between
the $B_c^+\to\jpsi a_1(1260)$ and 
$B_c^+\to\psi(2S)\pi^+$, $\psi(2S)\to\jpsi\pi^+\pi^-$ decays, as obtained from the simulation. 
    
Other systematic uncertainties are due to limited knowledge of the $B_c^+$ lifetime \cite{PDG} ($4\%$),
uncertainty in the simulation of charged tracking efficiency ($5\%$), trigger ($4\%$) and the kaon veto ($5\%$).
Summing all contributions in quadrature, 
the total systematic error on the branching fractions ratio amounts to 14\%.
As a result, we measure the branching fraction ratio
\begin{equation*}
\frac{\BR(\bcjppp)}{\BR(\bcjp)}=2.41\pm0.30\pm0.33,
\end{equation*}
where the first uncertainty is statistical and the second systematic. 

The obtained result can be compared to theoretical predictions; these
assume factorisation into $B_c^+\to\jpsi W^{+*}$ and $W^{+*}\to\pi^+[\pi^-\pi^+]$.
The contributions of strong interactions to $B_c^+\to\jpsi W^{+*}$ are included in  
form-factors which can be calculated in various approaches such as 
a non-relativistic quark model or sum rules.
The coupling of a single pion to a $W^{+*}$ is described by the pion decay constant.
The coupling of three pions to a $W^{+*}$ is measured in $\tau^-\to\nu_\tau\pi^-\pi^+\pi^-$ decays,
which are dominated by the $a_1(1260)$ resonance.
The prediction by Rakitin and Koshkarev, using the no-recoil approximation in $B_c^+\to\jpsi W^{+*}$, is
$\BR(\bcjppp)/\BR(\bcjp)= 1.5$ \cite{Rakitin:2009ya}.
Likhoded and Luchinsky used three different approaches to predict the form factors and obtained
$\BR(\bcjppp)/\BR(\bcjp)= 1.9, 2.0$ and $2.3$, respectively \cite{Likhoded:2009ib}.
Our result prefers the latter predictions. 
It is also consistent with 
$\BR(\budppp)/\BR(\budp)=2.00\pm0.25$ \cite{PDG},
which is mediated by similar decay mechanisms, and with a similiar ratio of phase-space factors.
Our result constitutes the first test of theoretical predictions for branching fractions of $B_c^+$ decays.

\section*{Acknowledgements}

We express our gratitude to our colleagues in the CERN accelerator
departments for the excellent performance of the LHC. We thank the
technical and administrative staff at CERN and at the LHCb institutes,
and acknowledge support from the National Agencies: CAPES, CNPq,
FAPERJ and FINEP (Brazil); CERN; NSFC (China); CNRS/IN2P3 (France);
BMBF, DFG, HGF and MPG (Germany); SFI (Ireland); INFN (Italy); FOM and
NWO (The Netherlands); SCSR (Poland); ANCS (Romania); MinES of Russia and
Rosatom (Russia); MICINN, XuntaGal and GENCAT (Spain); SNSF and SER
(Switzerland); NAS Ukraine (Ukraine); STFC (United Kingdom); NSF
(USA). We also acknowledge the support received from the ERC under FP7
and the Region Auvergne.

\bibliographystyle{LHCb}
\bibliography{main}
\end{document}